\documentclass[preprint2]{aastex}

\shorttitle{Optical -- Radio Variability}
\shortauthors{De Vries et al.}

\begin{document}
\title{Optical Properties of faint FIRST Variable Radio Sources}

\author{W. H. de Vries, R. H. Becker}
\affil{University of California, One Shields Ave, Davis, CA 95616}
\affil{Lawrence Livermore National Laboratory, L-413, Livermore, CA 94550}
\email{devries1@llnl.gov}

\author{R. L. White}
\affil{Space Telescope Science Institute, 3700 San Martin Drive,
Baltimore, MD 21218}

\and

\author{D. J. Helfand} \affil{Columbia Astrophysics Laboratory,
Columbia University, 550 West 120th Street, New York, NY 10027}

\begin{abstract}
  
  A sample of 123 radio sources that exhibit significant variations at
  1.4 GHz on a seven year base-line has been created using FIRST VLA
  B-array data from 1995 and 2002 on a strip at $\delta=0$ near the
  south Galactic cap. This sample spans the range of radio flux
  densities from $\sim2$ to 1000 mJy. It presents both in size and
  radio flux density range a unique starting point for variability
  studies of galaxies and quasars harboring lower luminosity Active
  Galactic Nuclei (AGN). We find, by comparing our variable and
  non-variable control samples to the Sloan Digital Sky Survey the
  following: 1) The quasar fraction of both the variable and
  non-variable samples declines as a function of declining radio flux
  density levels; 2) our variable sample contains a consistently
  higher fraction of quasars than the non-variable control sample,
  irrespective of radio flux; 3) the variable sources are almost twice
  as likely to be retrieved from the optical SDSS data than the
  non-variable ones; 4) based on relative numbers, we estimate that
  quasars are about five times more likely to harbor a variable radio
  source than are galaxies; and 5) there does not appear to be any
  significant optical color offset between the two samples, even
  though the suggestive trend for sources to be bluer when variable
  has been detected before and may be real. This leads us to conclude
  that both radio variability and radio flux density levels, in
  combination with accurate optical information, are important
  discriminators in the study of (radio) variability of galaxies. The
  latter start to dominate the source counts below $\sim 20$mJy. In
  any case, variability appears to be an intrinsic property of radio
  sources, and is not limited to quasars. Radio variability at low
  flux density levels may offer a unique tool in AGN unification
  studies.

\end{abstract}


\section{Introduction}

All is flux, nothing stays still; nothing endures but
change\footnote{Heraclitus (540-480 BC)}. This seems to apply
particularly well to extragalactic radio sources. Important advances
in our understanding of radio source intrinsic and extrinsic
variability mechanisms have been made by studying their amplitudes,
characteristic time-scales, and cross-correlations with observations
at other wavelengths. Radio variability can be classified in three
broad classes (Padrielli et al. 1987), based on their frequency
dependence: 1) variability only occurs at low frequencies ($\la
2$GHz), 2) variability occurs both at low and high ($\ga 5$GHz)
frequencies, but is not correlated in time, and 3) low and high
frequency variability, but the variations are correlated (although
they need not occur simultaneously). This classification also
conveniently separates the underlying variability mechanisms into
``most likely extrinsic'' (low frequency variations only), and ``most
likely intrinsic'' (correlated broadband variability).

The accepted model for the extrinsic variability of radio sources
is refractive interstellar scintillation (ISS), which predominantly
induces variations at low frequencies ($\la 1$GHz), with low
amplitudes ($\sim 2\%$), and with typical timescales of a few days
\citep{blandford86, rickett86}. Variations of this type have been
detected in multi-frequency studies (e.g., Mitchell et al. 1994; Lazio
et al. 2001).  Perhaps the strongest arguments for the external origin
of these variations are correlations with Galactic latitude (e.g.,
Gregorini et al. 1986; Gaensler et al. 2000), and an annual modulation
due to earth's orbital motion (e.g., Lainela \& Valtaoja 1993; Bondi
et al. 1994; 1996). Low-frequency variations with much longer
timescales ($\sim 10$ years), and/or larger amplitudes are thought
to be source intrinsic, however (e.g., R\'ys \& Machalski 1990 or,
more recently, Gaensler \& Hunstead 2000).

Broadband radio variability, in which the variations usually appear
first and more strongly at higher frequencies, are thought to be
intrinsic to the radio source. A model describing this behavior, in
which shocks propagate along the radio jets, has been proposed by
\citet{marscher85}. The model further implies that sources viewed
close to the line of sight (i.e., flat-spectrum quasars and BL Lacs)
should be more variable and have shorter variability timescales. For
objects very close to the line of sight, relativistic beaming will
become important, amplifying the variations and shortening the
timescales to $\la 1$ day (e.g., Lister 2001). Indeed, intraday
variability has been readily detected in samples of compact,
flat-spectrum sources. Some of the variability occurs on such short
timescales that the apparent source brightness temperature
(e.g., Gopal-Krishna et al. 1984) can be as high as $T\approx 10^{21}$K
(e.g., Quirrenbach et al. 2000), well beyond the inverse Compton limit
of $10^{12}$K (Kellermann \& Pauliny-Toth 1969). Doppler boosting
factors of up to a few hundred are needed to reconcile these numbers
(the apparent T is proportional to the Doppler factor to the third
power).  These Doppler factors are actually at the high end of the
observed range, and it has been suggested that these velocities are
pattern speeds that do not reflect the Lorentz factors of the jet
itself (e.g., Lister et al. 2001). Nonetheless, intraday variability
has been uniquely identified with the blazar class of objects
(e.g., Quirrenbach et al. 1992; Lister et al. 2001).

\begin{figure*}[th]
\epsscale{2.0}
\plotone{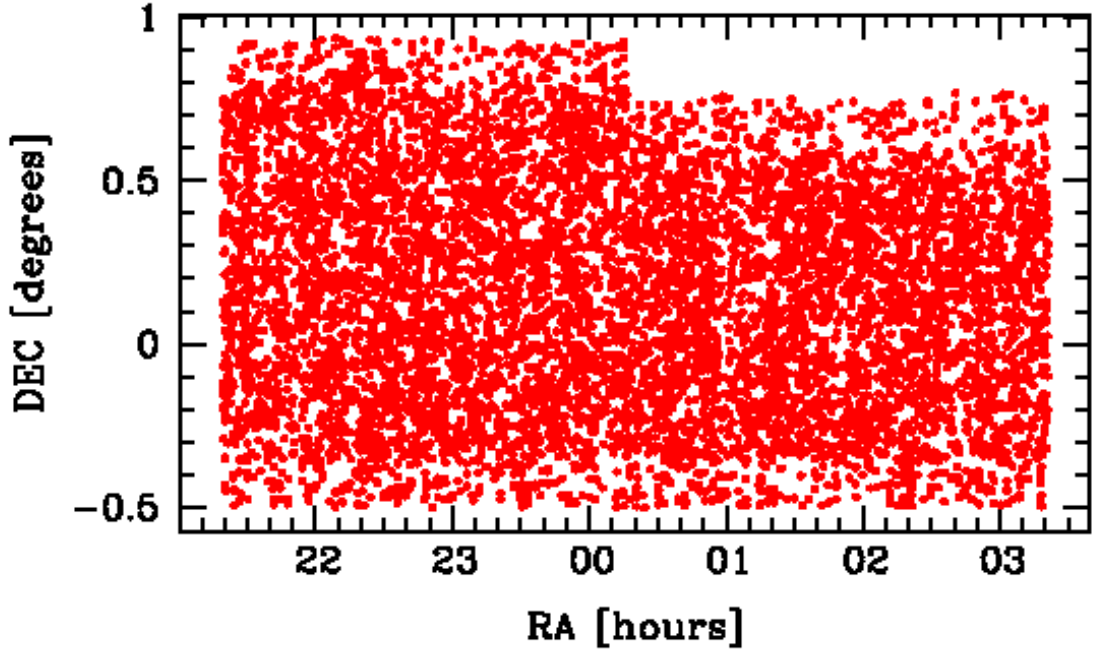}
\caption{FIRST Zero-Dec survey area. Note the non-uniform source
density close to the survey boundaries (due to the very different axis
scales this is not obvious for the RA edges).}
\label{area}
\end{figure*}

Comparatively little is known about (intrinsic) radio variability in
non-blazar sources at low flux density levels ($\sim 10$ mJy). Most of
the variability research has focused on either bright radio samples
(e.g., Gregorini et al. 1986, $S > 0.4$ Jy ; Lister et al. 2001, $S >
0.4$ Jy; Aller et al. 2003, $S > 1.3$ Jy), or small samples at much
lower flux levels (a few 10's of mJy -- e.g., R\'ys \& Machalski 1990;
Bondi et al. 1996; Riley et al. 1998). Recently, Carilli et al. (2003)
used deep multi-epoch VLA\footnote{The Very Large Array is a facility
  of the National Radio Astronomy Observatory (NRAO), which is
  operated by the Associated Universities, Inc., under cooperative
  agreement with the National Science Foundation.}  observations of
the Lockman Hole area to study variability at the 0.1 mJy level,
resulting in just 9 variable sources.

In this paper, we will use two epoch observations of the FIRST
Zero-Dec strip (Sect.~\ref{observations}) of about 9000 radio sources
to expand significantly the number of variable sources down to flux
density levels of $\sim 1$ mJy. Unlike much brighter radio samples,
the typical radio source population at these levels is dominated by
star-forming galaxies and steep-spectrum systems, with only a small
percentage of flat-spectrum AGN (e.g., Windhorst et al.  1999;
Richards et al. 1999; de Vries et al. 2002). With our sample,
therefore, we are in an excellent position to investigate the radio
variability properties of lower radio luminosity galaxies.

Interstellar scintillation is not expected to contribute significantly
to the variability in our sources, since our observations were carried
out at 1.4 GHz, near the minimum in expected ISS (e.g., Padrielli et
al.  1987; Mitchell et al. 1994). Furthermore, given our variability
criterion (see Sect.~\ref{sample}), only sources that vary by
significant amounts, especially toward the lower flux levels, are
included. These variation levels exceed the typical ISS variations of
a few percent by a large factor.

The paper is outlined as follows. In the next two sections, the radio
observations and sample selection are described. In Sect.~\ref{optid}
we discuss the optical properties of the variable sample, and compare
these against a non-variable control sample.

\section{Observations} \label{observations}

Our radio imaging data have been taken as part of the Faint Images of
the Radio Sky at Twenty centimeters survey (FIRST, see Becker et al.
1995 for a detailed description) conducted with the VLA in B-array
between 1993 and 2002.  The observing strategy of the FIRST survey
provides a limited capability to search for variability, although
since adjacent fields have significant overlap, most sources were
observed more than once. For the most part, data were collected in
strips of constant declination so adjacent fields in the east-west
direction are sensitive to variations on a time scale of 3
minutes. FIRST observing runs were usually separated by multiples of
24 hours so adjacent fields in the north-south direction are sensitive
to variations on timescales of 1 to 5 days.  This led, for instance,
to the discovery of a number of radio-variable stars
\citep{helfand99}.

A search for variability between the FIRST survey and the NRAO VLA Sky
Survey (NVSS, Condon et al. 1998) is also possible. This is, however,
complicated by the difference in angular resolution: the FIRST survey
can resolve out flux seen by the NVSS. This results in so many NVSS
sources appearing brighter than their FIRST counterparts that it is
necessary to restrict the analysis to sources which appear brighter in
FIRST (see Sect.~\ref{nvsscomp}), making a straightforward flux
comparison to test for variability difficult.

The best opportunity to search for long term variability has been
afford by our decision to re-observe the south Galactic cap (SGC)
FIRST Zero Dec strip (hereafter FZDB) during the summer of 2002 (the
initial set of observation were made in 1995, which will be referred
to as FZDA). The observations were made for the dual purpose of a
quality control test of the FIRST survey, as well as a search for
variability. The zero-dec strip was chosen for this purpose because it
has been observed by the Sloan Digital Sky Survey as well (SDSS, see
Stoughton et al. 2002 for the Early Data Release, and Abazajian et
al. 2003 for Data Release 1 updates). A total of 720 fields were
re-observed covering 120.2 square degrees and encompassing 9086 radio
objects in the FZDB catalog. The FZDB survey area is shown in
Fig.~\ref{area}.

\subsection{Matching FZDB with FIRST}

\begin{figure}[t]
\epsscale{1.0}
\plotone{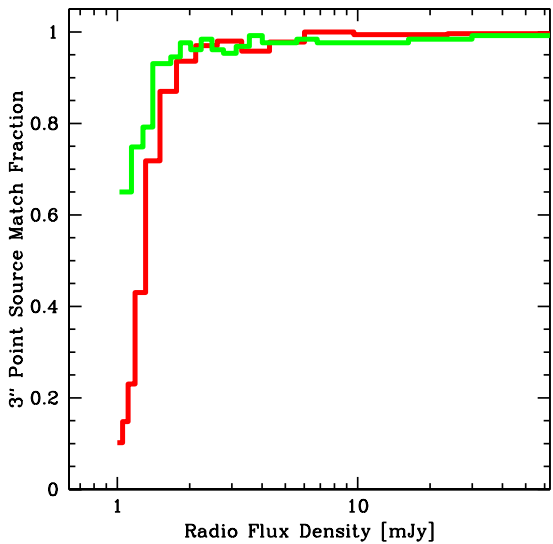}
\caption{Fractions of the FZDA and FZDB catalogs (green and red lines
respectively) that are present in the combined catalog. The histograms
are for point sources only, and the matching has been done to within
3\arcsec. Note the high fractions for sources with flux densities
exceeding 2 mJy (98.0\% on average). It is also apparent that some
fraction of FZDB sources below 2 mJy are spurious, more so than for
the FZDA catalog. Each bin represents 500 sources, so the closer
spacing toward lower flux density levels reflects the increasing
number of faint objects.}
\label{matchfrac}
\end{figure}

The FIRST catalog has a uniform sensitivity across the surveyed
zero-dec strip area, whereas the FZDB has not (Fig.~\ref{area}).  This
makes a direct comparison of their relative completeness limits and
source surface densities difficult.  With an average surface density
of $\sim90$ radio sources per square degree \citet{white97}, the FIRST
catalog should contain about 10\,800 objects within the nominal
zero-dec strip survey area. However, this number cannot be directly
compared to the FZDB count of 9086, due to the latter's non-uniform
survey coverage. The best way to compare the two catalogs is to create
an FZDA catalog (derived from FIRST) which has its source surface
density tapering off toward the survey edges, just like the FZDB
catalog. The simplest way of obtaining such a catalog is to list the
unique sources contained in FIRST around the FZDB catalog positions.
This makes sense, since we are interested in objects present in both
catalogs in the first place. The total number of sources in such an
FZDA catalog vary with the radius up to which the FZDB positions are
matched: 6605 sources within 3\arcsec, 7343 within 1\arcmin, and 11778
within 10\arcmin. We will use the first two lists to compare the
relative completeness of the FZDB catalog. The 6605 entry list
represents our matched catalog from which we will select our variable
sample (Sect.~\ref{varcrit}), and the 7343 constitutes our comparison
FZDA sample.

The combined catalog uses a matching radius of 3\arcsec, which is
about half of the FWHM of the B-array beam. This automatically
excludes the fraction of extended sources for which the source
detection program assigned different components and/or different
positions though. As a consequence, the number of 6605 is considerably
lower than the original FIRST surface density (even lower than
expected based on the non-uniform coverage).  To be consistent, we
exclude {\it all} extended components\footnote{We consider a source /
component extended if it has a peak flux less than 70\% of its
integrated flux. It should be noted that we are referring to
components, so, for instance, it is possible that a single extended
radio source can be resolved into a number of unresolved components,
each of which would end up individually in our combined sample; see
Sect~\ref{radiomorph}.}  from our combined sample. There are good
scientific reasons for this as well. Extended radio emission is most
likely due to lobe (radio hotspot) emission which cannot vary on the
short timescales we are sampling here ($7 / (1+z)$ years). Since we
are interested in shorter timescales, we will have to study those
sources in which the (feeding) processes close to the AGN itself
vary. These sources are consequently very small, and unresolved by our
imaging data (with its resolution of $\sim5$ arcseconds). The
resulting reduced sample sizes have been listed in
Table~\ref{samples}.

\subsubsection{Sources unique to one catalog - Transient phenomena}

Figure~\ref{matchfrac} illustrates how well the FZDA and FZDB compare.
At the higher flux density levels we expect to find almost all of the
point sources in the combined catalog. This is the case for sources
with flux densities exceeding 2 mJy. For both the FZDA and FZDB
catalogs, over 98\% of those sources end up in the combined catalog.
None of the remaining catalog-unique entries turned out to be real on
close inspection: they were either resolved sources for which a
particular component was assigned a differing position between FZDA
and FDZB (separated by more than 3\arcsec), or the catalog entry
turned out to be an imaging artifact that made it into one but not
both catalogs (e.g., ringing around very bright objects).

The situation is quite different for fainter sources, however. An
increasing fraction of faint objects are only present in either
catalog, which indicates a systematic effect rather than genuine
source variability. Disentangling the truly variable sources from the
random noise at these flux density levels can only be achieved by
obtaining deeper surveys. Indeed, none of these faint sources end up
meeting our variability criterion (Sect.~\ref{varcrit}), which is
based on the survey sensitivity level. All the variable sources that
{\it do} meet the criterion have flux densities $>2$mJy (see
Fig.~\ref{vp2MED}), and are in the well-matched part of
Fig.~\ref{matchfrac}. 

So, in summary, transient phenomena (e.g., Gamma Ray Burst afterglows)
have not been found for flux density limits exceeding 2 mJy. Fainter
events may be present, but we do not have any way of assessing their
reality.

\section{Variable sample selection}\label{sample}

Since we are interested in radio source variability, we begin by
defining our Variability Ratio (hereafter VR) as:

\begin{equation}
\mbox{VR}(i) = \frac{S(i)_{\mbox{\footnotesize B}}}{S(i)_{\mbox{\footnotesize A}}}
\label{vreqn}
\end{equation}

\noindent for the $i^{\mbox{th}}$ source in our catalog. The indices A
and B denote the measurement epochs 1995 and 2002.  Because the source
brightness does not enter into this equation, we cannot use this ratio
at face value: toward the lower flux density levels the (sky
background) noise will account for a significant fraction of the
``variability'' in such sources. We therefore have to weight the VR
value of each source individually based on its Signal-to-Noise Ratio
(S/N) and local noise values (see next section).

\begin{figure}[t]
\epsscale{1.0}
\plotone{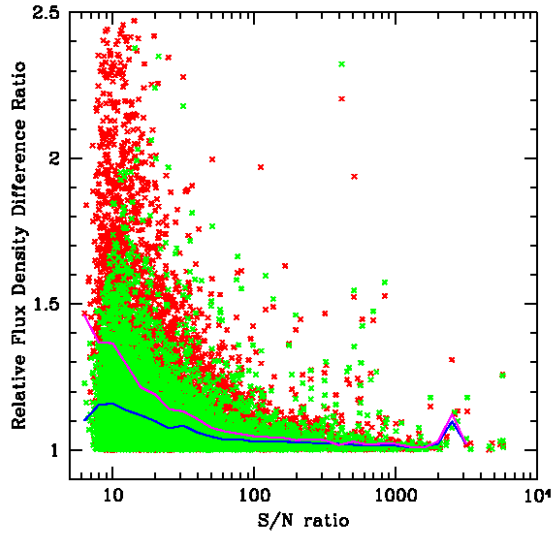}
\caption{Relative flux density ratios for our point source sample as
function of S/N. The flux density ratio is defined here as larger flux
density divided by smaller flux, irrespective of the epoch (hence it
is always $>1$). The red points are for the integrated flux density
values, and the green points for just the fitted peak values. As can
be seen by the overplotted medians (purple for integrated, blue for
peak flux densities), the former has a larger scatter, especially
toward lower flux density levels.}
\label{relvar}
\end{figure}

Since we are only considering unresolved sources, ideally their
(fitted) peak flux density and integrated flux density should be the
same. This is, however, not the case. In fact, the integrated flux
density values are found to have larger intrinsic scatter than the
fitted peak flux values. Both the peak and integrated flux density are
based on Gaussian fits to the actual data, but unlike the peak fit,
the integrated value is also based on fits to the major and minor axes
of the object.  Noise in the measurement of the axes therefore
increases the noise in the integrated flux density compared to the
peak value.  In Fig.~\ref{relvar} we have plotted the relative flux
density difference ratio (defined as the larger flux density divided
by the smaller flux density) for both the source integrated and peak
flux density values.

It is clear from this figure that the scatter in the peak flux
measurements (the solid blue median line) is considerably less than
that of the integrated flux density ratios (as indicated by the solid
purple median line). To minimize the number of variable false
positives, we only consider the source peak flux density values.

\subsection{Variability criterion}\label{varcrit}

\begin{figure}[t]
\epsscale{1.0}
\plotone{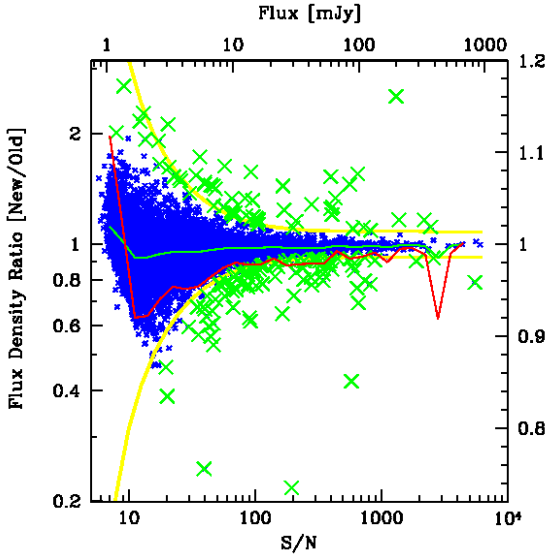}
\caption{Flux density variations for unresolved sources between the
two epoch as function of signal-to-noise. The peak flux density value
of the most recent epoch has been divided by the 1995 data.  The
yellow lines represent the 4$\sigma$ variation limits, as given by
Eqn.~\ref{fluxerror3}. The highlighted green crosses are sources that
meet our variability criterion, the blue crosses are considered
non-variable. The solid green line marks the local median variation
value, and serves as an indication of the relative calibration between
the two epochs. This line has been replotted for better clarity on a
linear scale (indicated on the right vertical axis) as a solid red
line.}
\label{varplot2}
\end{figure}

\begin{figure}[t]
\epsscale{1.0}
\plotone{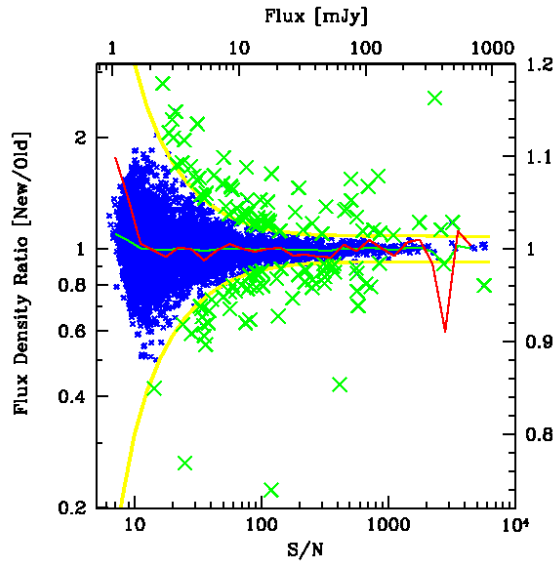}
\caption{The same plot as Fig.~\ref{varplot2}, but with the adjusted
FZDB flux densities: $1.0116\times(S_B + 0.090\ \mbox{mJy})$. This
straightens out the median variability curve and evens out the number
of positive and negative variable sources.}
\label{vp2MED}
\end{figure}

Following \citet{rengelink97}, the relative flux density errors can be written
as:

\begin{equation}
\frac{\sigma_S}{S} = \left( C_1^2 + C_2^2 \left( \frac{\sigma_{\mbox{rms}}}{S} 
\right)^2 \right)^{\frac{1}{2}}
\label{fluxerror}
\end{equation}

\noindent This equation reflects the two components of the measurement
error, with $C_1$ due to a constant systematic error, and $C_2$
dependent on the S/N ratio. This expression can be used to estimate
the significance of radio source variation, as defined by the quotient
of the new and old flux densities. If one defines the flux density and
noise values in the first epoch as $S_A$,$\sigma_A$, and in the second
as $S_B$,$\sigma_B$, the 4$\sigma$ variability envelopes are then
given by:

\begin{equation}
\mbox{VR}_4 = \frac{S_B\left( 1 \pm 4\sqrt{C_1^2+C_2^2(\frac{\sigma_B}{S_B})^2} \right)}
{S_A\left( 1 \mp 4\sqrt{C_1^2+C_2^2(\frac{\sigma_A}{S_A})^2} \right)}
\label{fluxerror2}
\end{equation}

\noindent which can be further reduced by substituting $S_A = S_B =
S_{\mbox{true}}$, $\sigma_A = \sigma_B = 0.15$ mJy, $C_1 = 0.01$, and
$C_2 = 1.3$. The $\sigma$ values have been measured to be on average
0.150 and 0.148 mJy for the 1995 (A) and 2002 (B) epochs,
respectively. The $C_2$ value has been chosen to match the value of
\citet{rengelink97} and \citet{devries02}, and seems fairly robust
across different instruments and survey setups. The $C_1 = 0.01$ value
(which is lower than the $C_1 = 0.04 $ of the latter two papers), is
set by the high S/N end of the source distribution. Its adopted value
is a good match between including / excluding high S/N variable
sources: the fraction of variable sources as function of S/N is kept
acceptably flat (see Figs.~\ref{varplot2} and \ref{vp2MED}, with S/N
defined as $(S_A+S_B)/(\sigma_A+\sigma_B)$).

The upper and lower $4\sigma$ envelopes can now be expressed (as a
function of S/N) as:

\begin{eqnarray}
\mbox{VR}_4^{\mbox{\footnotesize upper}} = \frac{1 + 4\chi}{1 - 4\chi}\nonumber\\
\mbox{VR}_4^{\mbox{\footnotesize lower}} = \left( \mbox{VR}_4^{\mbox{\footnotesize upper}} \right)^{-1}\nonumber\\
\mbox{with: } \chi = \sqrt{0.01^2+\left( \frac{1.3}{\mbox{S/N}}\right)^2}
\label{fluxerror3}
\end{eqnarray}

Equation~\ref{fluxerror3} forms the basis of our variability
criterion. Any source having a VR exceeding either $4\sigma$ envelope
is considered genuinely variable. This yields 146 variable sources, of
which 58 are brighter in the 2002 epoch, and 88 are fainter
(Fig.~\ref{varplot2}).

The solid red curve in Fig.~\ref{varplot2} represents the median flux
density ratio as function of S/N, plotted on a linear scale to improve
clarity (as opposed to the logarithmic equivalent in green). As can be
gathered from this curve, the medians quickly fall below 1 as the S/N
decreases, with a maximum deviation of 8\% at S/N $\sim10$. Most of
the offsets are within $\sim2$\%, however.  Even though the offsets
are small, there is a systematic trend which is affecting our
variability sample selection, something we wish to avoid. If
uncorrected, it implies that fainter sources are more likely to fade
between 1995 and 2002 than brighter sources, something which would be
hard to explain in any variability scenario. We have two parameters to
play with: a small (additive) zero-point correction which will
straighten out the median curve, and a multiplicative sensitivity
correction which will raise or lower the curve along the y-axis. The
best correction (with the resulting median curve having the smallest
variations along the $\mbox{VR$_c$}=1$ line) is obtained by adjusting
the FZDB flux densities by:

\begin{equation}
\mbox{VR$_c$}(i) = \frac{1.0116\times\left( S(i)_{\mbox{\footnotesize B}} + 0.090\  \mbox{mJy}\right)}
{S(i)_{\mbox{\footnotesize A}}}
\label{vreqn2}
\end{equation}

\noindent which amounts to a 90 $\mu$Jy zero-point offset, and a
1.16\% sensitivity correction. The 90 $\mu$Jy offset is small compared
to both the 150 $\mu$Jy rms noise value and the 250 $\mu$Jy CLEAN bias
(a flux density correction applied per beam for the FIRST survey,
see White et al.  1997). And the 1.16\% factor is well within the
nominal $\sim5\%$ systematic uncertainty in the flux density scale.

After applying these corrections (Fig.~\ref{vp2MED}), we end up with
128 candidate variable sources / components (corresponding to 2.0\% of
the sample), listed in Table~\ref{VarSample}, of which 70 are brighter
in the 2002 epoch, and 58 fainter.

\subsection{NVSS comparison}\label{nvsscomp}

As a reality check, we compared the FZDB catalog against the NVSS
catalog to see whether the same sources are deemed variable, using
basically an identical variability criterion. There are, however, a
few differences. The NVSS survey has a typical background noise of
0.45 mJy (Condon et al. 1998), instead of the 0.15 mJy for the FIRST
survey. Another complicating factor is the resolution difference
between the NVSS and FZDB. Sources that are not resolved by the
$45\arcsec$ NVSS beam are often resolved by the FZDB, with each
component only a fraction of the NVSS total flux density. We tried to
minimize this effect by first excluding resolved NVSS sources outright
from the correlation, and secondly using the FZDB integrated (instead
of peak) flux density measurements. Furthermore, we used a very
restrictive $5\arcsec$ matching radius in another attempt to just
match unresolved sources both in the FZDB and NVSS. The resulting
matched sample contains 2201 sources.

Of these 2201 sources, 10 are brighter at the FZDB epoch, and 91 are
fainter. This is after applying a similar FZDB flux density correction
as in the previous section (albeit with slightly different constants).
Clearly, there is still some flux density being resolved out by the
longer base-line FZDB observations compared to the NVSS, which results
in quite a few bogus variable sources.

This implies we cannot really compare the FZDB-NVSS faint variable
sources to Table~\ref{VarSample}, since an unknown fraction of them
will be spurious. The 10 FZDB-NVSS bright variable sources on the
other hand, are not be affected by this. Indeed, 9 out of the 10
sources are recovered in Table~\ref{VarSample} (separately footnoted),
with the sole ``newly'' identified variable source FIRST
J222646$+$005210. Upon closer inspection, this source appears to be
stable between FZDA and FZDB ($653\pm1$ mJy for both epochs). Its NVSS
flux density of 615 mJy is just within the quoted $\sim5\%$ systematic
flux density uncertainty between the two surveys. Based on this, we
therefore conclude that our variability criterion is robust, and
results in a sample of truly variable sources. This does tacitly
assume that the cataloged radio flux densities are always
correct. However, as the next section will show, this is actually not
valid for some cases.

\section{Radio morphology}\label{radiomorph}

Our variable sample actually contains 128 variable, unresolved, {\it
components}. While most of these component entries turn out to be
single (unresolved) radio pointsources, some of them are part of a
multi-component radio source. The varying part is then presumably
associated with the AGN / radio-core position. However, there may
still be some cases where the FZDA and FZDB cataloging software
divided up radio flux differently between adjacent components,
introducing artificial variability. It is therefore important to check
all the multi-component radio sources in our sample (28)
individually.

In order to characterize the {\it overall} radio properties of our
sample, we start by defining 6 morphological categories: 1) Single
pointsource (PS in Column 11 of Table~\ref{samples}); 2) Pointsource
with an elongation, or linear feature (CJ, ``core-jet''); 3) Variable
core with two non-variable lobes (CL); 4) Core embedded in diffuse
halo (CH); 5) Complex, multi-component structure (CX); and 6) Possible
hotspot variability (HS).

Each of the sources in Table~\ref{samples} have been categorized
accordingly. The relative number break-down of the sample is: 1) 100
(78.1\%); 2) 14 (10.9\%); 3) 9 (7.0\%); 4) 1 (0.8\%); 5) 1 (0.8\%);
and 6) 3 (2.3\%). Except for the last two classes (containing 4
sources, or 3.1\% of the total sample), all of the variability is
consistent with it being due to the AGN / core component. We will
discuss a few selected sources in more detail below. Radio maps of all
the extended variable radio sources can be found in
Fig.~\ref{radiomaps}.

\medskip
\noindent{\bf J000257$-$002447 (CH)} -- This source contains a variable
pointsource component embedded in an extended radio halo. The radio
surface brightness of this halo remains constant between the FZDA and
FZDB epochs, whereas the peak flux density of the central pointsource
varies by more than 10\%.

\noindent{\bf J001800$+$000313 (HS)} -- The North-Eastern lobe of this
double radio source is slightly resolved, just enough so that it did
not make it into our variable component list. As it turns out, both
lobes / hotspots appear to be variable. The total flux density for the
radio source increased by 13.6\%\ between FZDA and FZDB, from 62.14 to
70.56 mJy. Unfortunately, two unrelated sources close to this source
also ``varied'' by $\sim10\%$, casting serious doubt on its reality.
The source is very close to the edge of its particular survey field,
which may be the reason for the flux density discrepancy.  It is taken
out of further consideration, and has been placed separately in
Table~\ref{samples}.

\noindent{\bf J003246$-$001917 (HS)} -- Upon close inspection of both
the FZDA and FZDB catalogs, we have to come to the conclusion that
this source is not variable. The problems stems from the fact that
FZDA lists 5 components for this source, and FZDB only 4. This leads
to quite different flux density assignments to the North-Western lobe.
The source is taken out of further consideration.

\noindent{\bf J012213$-$001801 (CL)} -- Both the Eastern and Western
lobes are constant to within $\sim 1\%$ between FZDA and FZDB. Flux
density values for the bright Eastern lobe are 123.84 and 122.61 mJy,
a variation of 1.0\%. The core, on the other hand, brightened over the
same time period from 348.28 to 404.79 mJy.

\noindent{\bf J020234$+$000301 (CL)} -- The lobes do not vary beyond
the 1\%\ level. The core, however, fades from its FZDA flux of 44.01
mJy to 35.92 (an 18.4\% decrease).

\noindent{\bf J021840$-$001515 (CX)} -- This is the only source in our
sample with a hard to categorize radio morphology. The marked
component position (Fig.~\ref{radiomaps}) coincides with a $z=1.171$
quasar. The two components immediately to the North do not have
optical counterparts, though there appears to be a faint object in the
SDSS image that is situated in between the lobes. All of the radio
components fade between FZDA and FZDB by on average 16\%. A nearby,
unrelated, radio source observed in the same field remains constant,
however.  The component associated with the quasar varies the most:
from 14.12 mJy down to 10.92 mJy (a decline of 22.7\%). At the moment
it is not clear to us how the various components relate to each other,
if at all.

\noindent{\bf J021202$-$002750 (CL)} -- The core component faded by
at least 4 mJy ($\sim8\%$) relative to the lobes.

\noindent{\bf J212058$+$000612 (HS)} -- As is the case for
J003246$-$001917, this source is not variable. The South-Eastern lobe
is constant, and the perceived variability in the North-Western lobe is
due to differing FZDA and FZDB component flux density assignments. This 
source is also taken out of consideration.

\noindent{\bf J212955$+$000758 (CL)} -- The ``variability'' is due to
the FZDA having 3 components, whereas FZDB lists 4. The summed fluxes
do not vary.

\noindent{\bf J220017$-$000133 (CL)} -- The core component in this
source brightened (by 20\%), so we are inclined to regard this source
as variable even though the number of fitted components is different
between the FZDA and FZDB. This faint extra component in the FZDA
catalog (just to the North of the core) does not appear to contribute
anything to the total flux of the core in the FZDB catalog (where this
component is absent), based on the fitting parameters. 

\noindent{\bf J222729$+$000522 (CL)} -- The core variability is 17.2\%
(relative to the lobes), with the flux declining from 97.51 mJy to
80.69 mJy.

\noindent{\bf J235050$-$002848 (CL} -- The core flux increased,
whereas the lobes remained more or less constant.

\noindent{\bf J235828$+$003934 (CL} -- All of the components faded by
about the same amount, casting serious doubt on the reality of core
variability. We have taken this source out of further consideration.

So, after close inspection, we found all 3 potential hotspot variable
sources (HS) and 2 core-lobe (CL) sources to be non-variable. They
have been excluded from our subsequent studies, but are left in
Table~\ref{samples} and Fig.~\ref{radiomaps} as reminders of how
careful one has to be using automated selection criteria.

\section{Optical identifications - Detection rates}\label{optid}

We next investigate the possible connection between optical properties
and the presence of radio variability.  Variability correlations in
the optical and radio have been proposed before (e.g., Tornikoski et
al. 1994; Hanski et al. 2002), with the highly beamed blazars as the
clearest examples of sources which exhibit variability across the
electromagnetic spectrum (e.g., Bregman et al. 1990). The first step
would be to correlate our variable sample with known optical catalogs,
and compare the results against a sample of non-variable radio
sources. We therefore created 12 sets of 123 non-variable sources
(i.e., each control sample has the same number of elements as the
variable sample). To ensure non-variability, we only selected sources
inside $1\sigma$ variation curves (as opposed to sources {\it outside}
$4\sigma$ variations, Fig.~\ref{vp2MED}), while matching the radio
flux density distribution function of the variable sample. This was
done by matching the number of non-variable sources to the amount of
variable sources per S/N bin (binned logarithmically from 1.0 to 3.4,
with binsize 0.4 dex).  Given that there are only 1922 sources (out of
our initial sample of 6605) that are less variable than these
$1\sigma$ limits, 12 sets of control samples are about as many as one
can create without starting to have a significant fraction of sources
shared between them.

All of our FZDA/FZDB survey area is covered by the Automatic Plate
Measuring (APM) machine catalog. The APM facility (in Cambridge, UK)
lists identifications and positions based on scanned UK and POSS~{II}
Schmidt plates, covering currently more than 15,\ 000 deg$^2$ of sky.

\subsection{APM matches}\label{apmid}

\begin{figure}[t]
\epsscale{1.0} 
\plotone{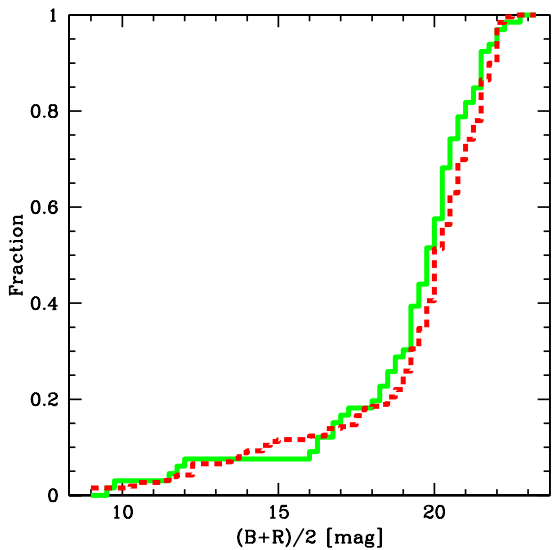}
\caption{Cumulative magnitude distribution of radio sources which are
identified by the APM (66 variable sources in total). The solid green
line is for our variable sample, the dashed red line represents the
distribution for the combined 12 non-variable (less than $1\sigma$)
control samples (262 sources in total). Both distributions are
identical, based on a KS-test (using 0.25 magnitude bins, for a total
of 56).}
\label{apmids}
\end{figure}

Our sample of 123 variable sources has been matched against the APM
catalogs with a $3\arcsec$ matching radius. While this does
discriminate against possible off-nuclear radio sources, it is
necessary to limit the number of chance matches, given the density of
objects in the APM catalog. The matching results are listed in
Table~\ref{apmmatch}. What is immediately striking is the difference
in detection rates between the variable and non-variable
samples. About half of the variable sources are retrieved from the APM
catalogs, but only 1 in 6 of the non-variable sources have optical
counterparts. The rms variation between the 12 control samples is
2.8\%, resulting in a $\sim 13\sigma$ detection rate difference.
Clearly, variable sources are easier to detect not only because on
average they are brighter, but also due to a higher fraction of
quasars\footnote{without spectroscopic confirmation, most of these
objects should be considered quasar candidates.}
(see Sect.~\ref{sdssmorph}).  The brightness difference is not large;
actually a cumulative distribution plot against mean magnitude
(Fig.~\ref{apmids}) shows how comparable their distributions
are. Based on a KS-test, these two distributions do not differ to any
level of significance.

Fortunately, a significant fraction ($\sim 60\%$) of the survey is
covered by the SDSS Data Release 1 (DR1), so we can use these deeper
and more uniform imaging data to investigate this further.

\subsection{SDSS} \label{sdssmorph}

The SDSS data also show a higher fraction of optical identifications
for variable sources. Since the latter data are deeper than the
photographic plates the APM catalogs are based on, we identify 82\% of
the variable sources, and on average 38\% of the non-variable ones
(Table~\ref{sdssmatch}). These values are up from the APM values of
54\% and 18\%, respectively. As mentioned earlier, one way of
explaining the higher match rate for the variable sample is that it
may contain a higher fraction of quasars.  Since these objects are
typically unresolved, they are easier to detect at any given magnitude
than resolved galaxies (of the same integrated
magnitude). Table~\ref{sdssmatch} lists the morphological breakdown of
each sample, based on the SDSS classification. With a relative stellar
fraction of 53\%, the variable sample contains significantly more
quasar candidates than the non-variable control sample (on average
19\%).

The results of Table~\ref{sdssmatch} can also be used to calculate how
much more quasars are likely to be variable compared to galaxies. This
is a function of radio flux density (Fig.~\ref{morphRF}), but the low
number of sources (per bin) only allow for an overall ratio.  We
assume the non-variable sample numbers (19\% of the radio population
are quasars, 81\% are galaxies) are representative of the general
population. These numbers compare very well to the ones quoted by
\citet{ivezic02} for a much larger sample of SDSS sources detected
with FIRST (83\% of those are galaxies). Now, if we introduce the
terms quasar variability rate QVR and galaxy variability rate GVR
(both defined as the fraction of quasars / galaxies that are
variable), and an initial sample size of $P$, we get:

\begin{eqnarray}
0.19 \times P \times \mbox{QVR} \equiv \mbox{\small \# of quasars in var. sample}\nonumber \\
0.81 \times P \times \mbox{GVR} \equiv \mbox{\small \# of galaxies in var. sample}
\end{eqnarray}

\noindent This means that the variable sample will contain in total
$(0.19 \times P \times \mbox{QVR} + 0.81 \times P \times \mbox{GVR})$
variable sources. We know the relative quasar and galaxy fractions
to be 53\% and 47\%. This leads to (after dividing out $P$):

\begin{eqnarray}
0.19 \times \mbox{QVR} = \nonumber \\
 0.53 \times (0.19 \times \mbox{QVR} + 0.81 \times \mbox{GVR}) \Leftrightarrow \nonumber \\
\mbox{QVR} = \frac{0.53 \times 0.81}{0.47 \times 0.19}\  \mbox{GVR} \nonumber \\
= 4.8\  \mbox{GVR}
\end{eqnarray}

\noindent If one assumes a $\sim 6\%$ uncertainty in the relative
fractions (Table~\ref{sdssmatch}), this relative value ranges between
3.4 and 7.5.  In other words, not surprisingly, quasars are on average
5 times as variable (in the radio) than are galaxies.

\subsection{Classification as function of radio flux density}

\begin{figure}[t]
\epsscale{1.0} 
\plotone{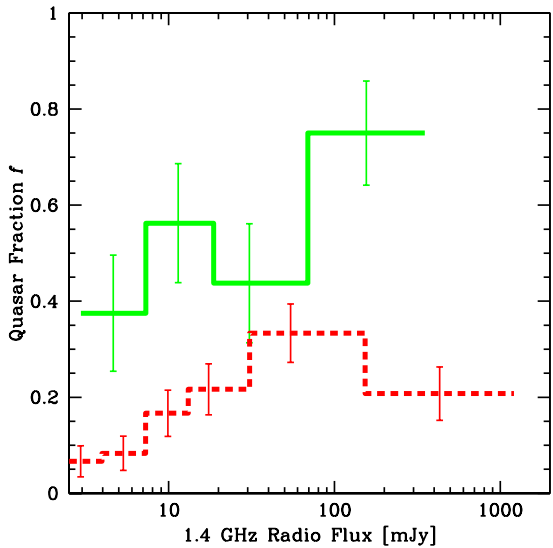}
\caption{Quasar population fraction as a function of radio flux density.  The
solid green line denotes the quasar fraction for our sample (64
objects in total, 16 per bin), and the dashed red line indicates the
fraction for our non-variable control sample (341 sources in total,
60 per bin). The error-bars are given by the formal binomial distribution
error: $\sigma(f)=(f(1-f)/N)^{0.5}$.}
\label{morphRF}
\end{figure}

The radio source population changes as a function of radio flux
density (e.g., Windhorst et al. 1999; Richards et al. 1999; de Vries
et al. 2002), from mainly AGN-dominated sources at higher flux density
levels down to starburst dominated sources at the sub-mJy level. This
will affect our variable sample as well, and we expect to see a
decline of the quasar fraction as function of lower flux density
limits. This is nicely illustrated in Fig.~\ref{morphRF}, which plots
the relative quasar fraction with flux density. The solid green line
represents our variable sample, and the dashed red line is the mean of
our 12 control samples.  The bins have a constant number of members
(except for the last bin in the control sample), hence the variable
spacing. The binning of the control samples has been adjusted to
approximately match the binning of the actual sample (4 bins of 16
each for the variable sample, and 5 bins of 60 plus 1 bin of 41
members for the control samples).

While it is not known how much uncertainty is present in the SDSS
classification (and hence quasar fraction) toward the faintest
magnitudes (see Sect.~\ref{clasrel} below), it is apparent from
Fig.~\ref{morphRF} that the quasar fractions for the variable and
non-variable samples differ significantly (the formal fraction errors
are 5\% for the control sample, and 12\% for the variable
sample). This leads us to two robust conclusions based on this figure:
1) a radio variable sample will contain more quasars than a
non-variable sample, irrespective of radio flux density, and 2) the
relative contribution of quasars to the sample declines as a function
of declining flux density: below $\sim 20$mJy, galaxies account for
the majority of the variable sources.

\begin{figure}[t]
\epsscale{1.0} 
\plotone{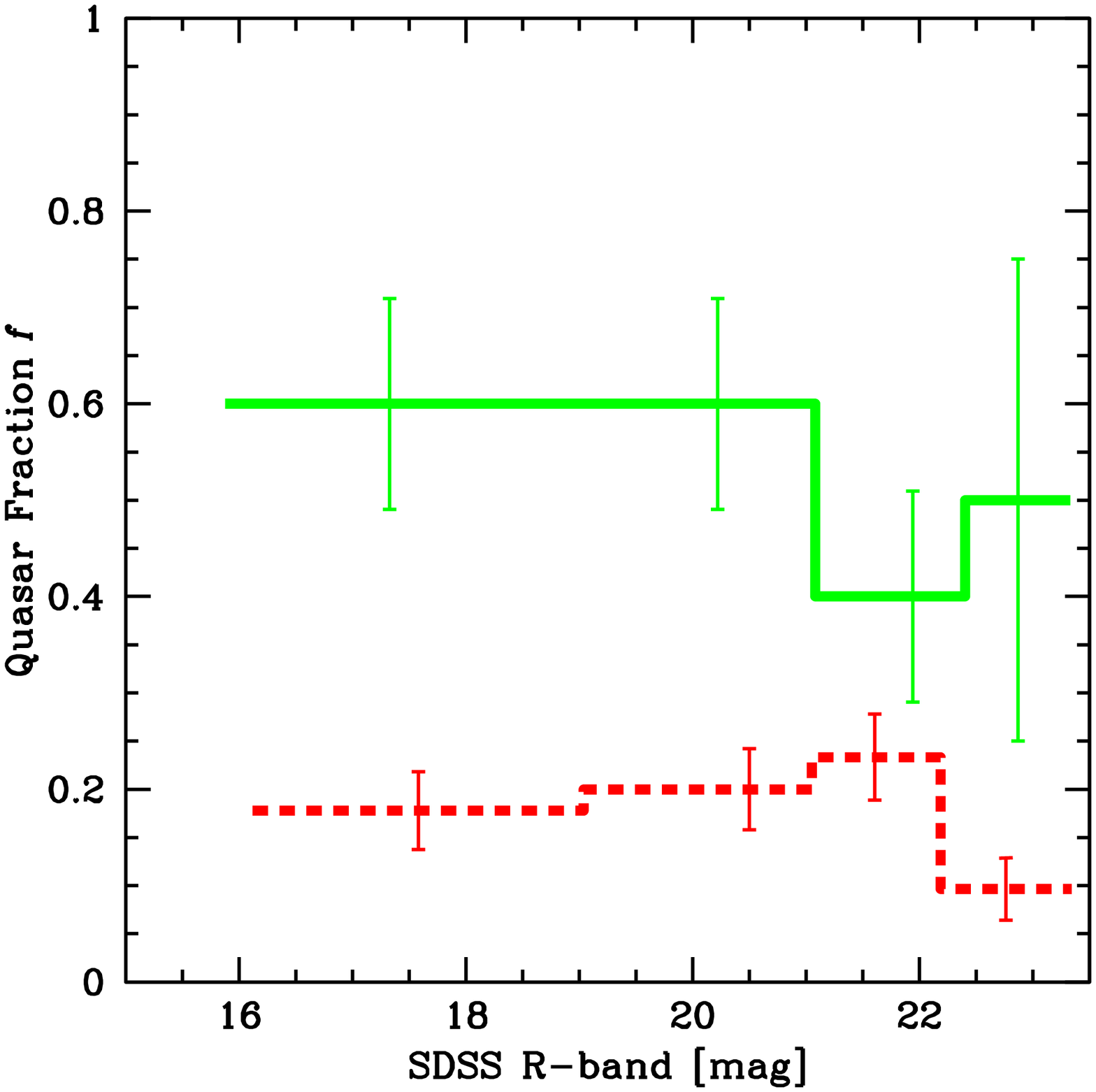}
\caption{Quasar population fraction as function of R-band magnitude.
Like in Fig.~\ref{morphRF}, the solid green line denotes the quasar
fraction for our variable sample, and the dashed red line is for the
non-variable control sample. We applied slightly larger bins here (20
and 90 objects for the variable and control samples).  Note that most
of the excess quasars in the variable sample are toward the brighter
magnitudes (R$<$21), and can be considered reliably
classified. Error-bars as defined in Fig.~\ref{morphRF}.}
\label{morphMAG}
\end{figure}

\subsubsection{Classification reliability}\label{clasrel}

The last section critically hinges on the reliability of the SDSS
galaxy versus quasar classification scheme. If for one reason, most of
the excess quasar classifications in our variable sample were made
toward the very faint end of the magnitude range, one might become
suspicious about the reality of the effect. A direct test is plotting
the quasar fraction as function of optical (R-band in this case)
magnitude, instead of radio flux density as in Fig.~\ref{morphRF}.
This has been done in Fig.~\ref{morphMAG}, with again our variable
sample in green and the control sample in red. It is reassuring to see
that most of the excess quasar identifications in the variable sample
are made for magnitudes brighter than $R \sim 21$. Stoughton et
al. (2002) estimate a 95\% confidence level for classifications up to
this magnitude.  It becomes much less certain for fainter sources.
Without doubt some of the current classifications are wrong, but there
is no indication that our consistently higher quasar fraction in the
variable sample, both as function of radio flux density and optical
magnitude, is significantly affected by this.

\subsection{Optical colors}

\begin{figure*}[t]
\epsscale{2.0} 
\plottwo{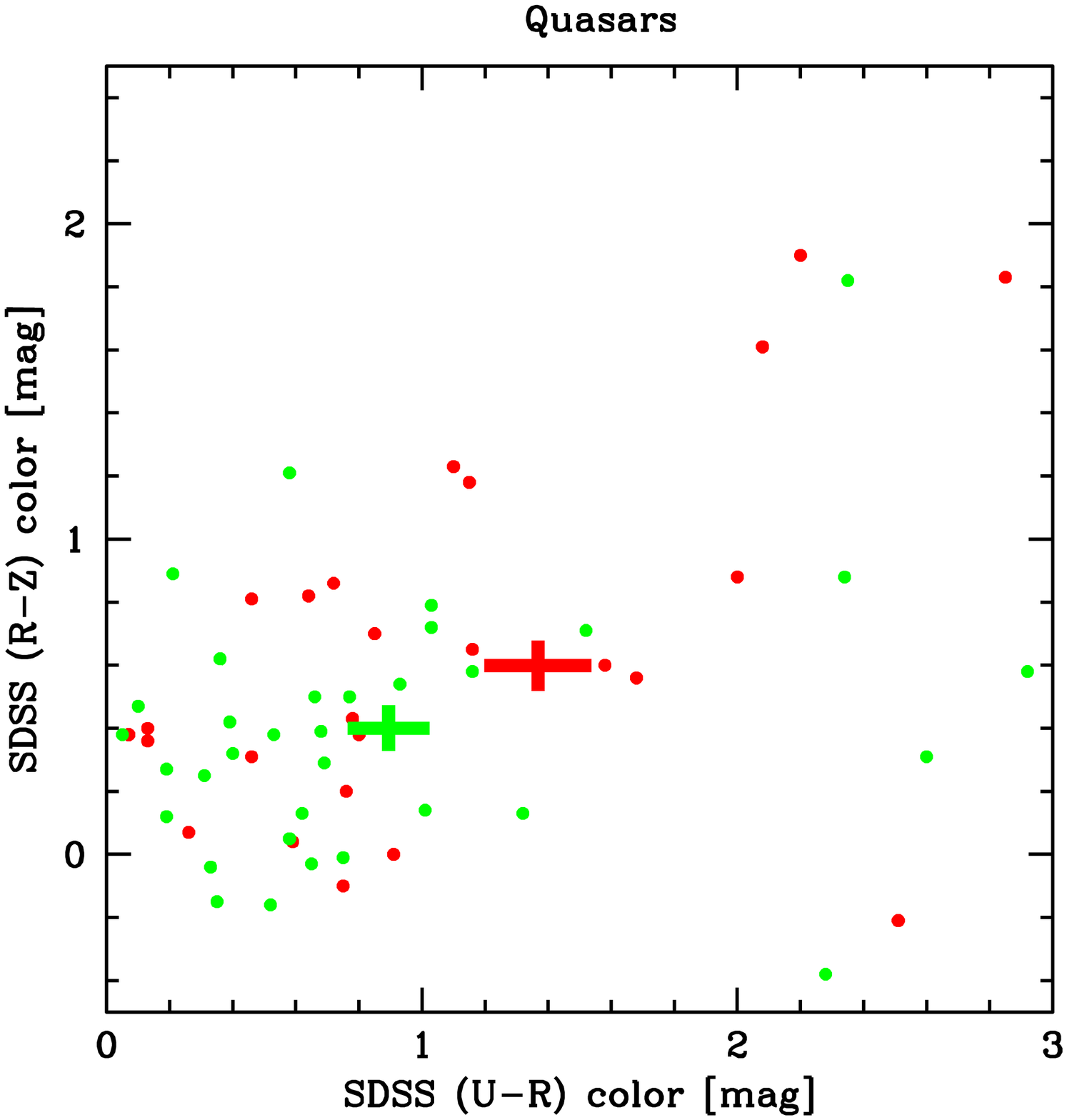}{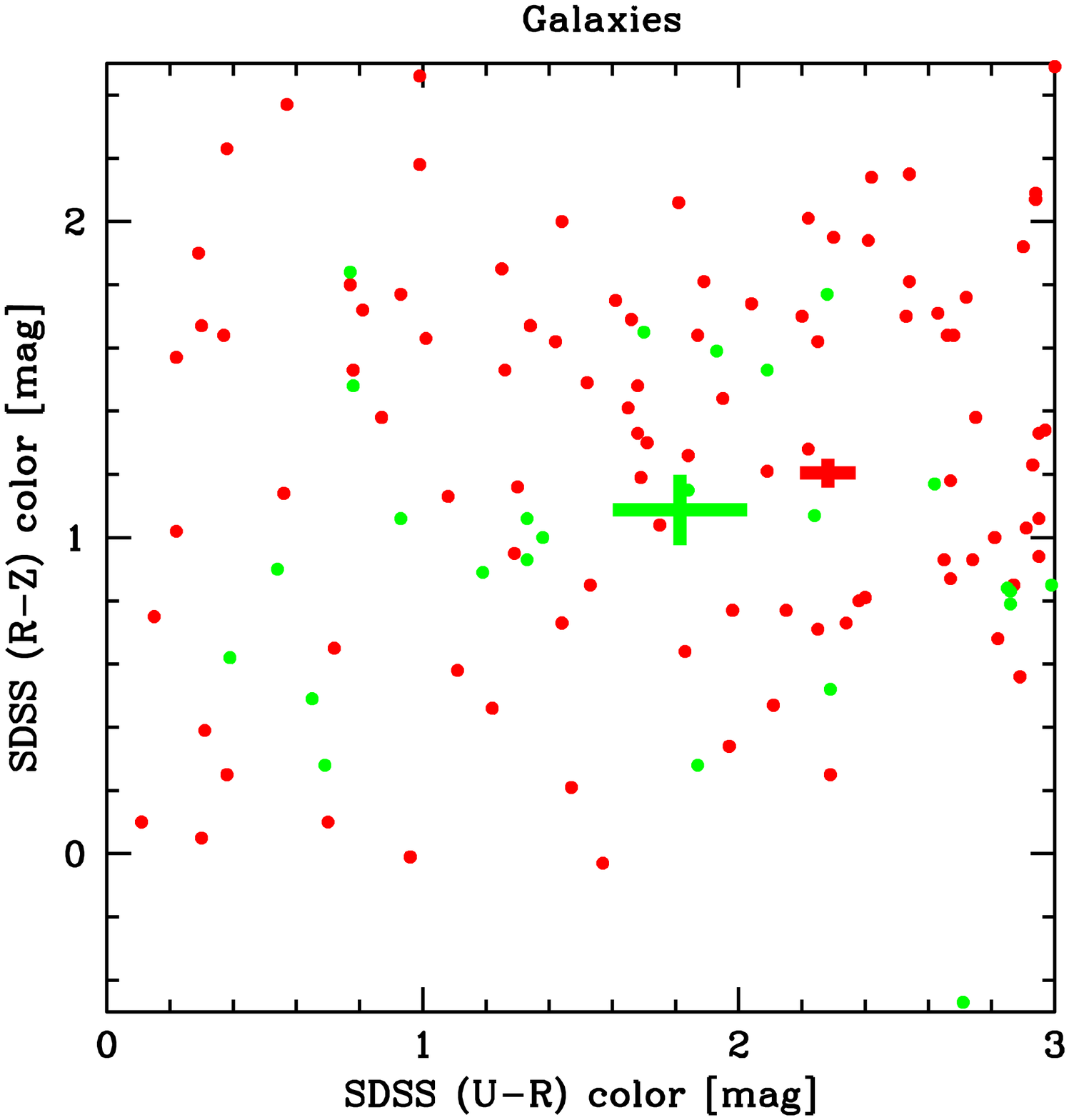}
\caption{SDSS (r$-$z) versus (u$-$r) color plots. The variable sample
is represented in green, with quasars and galaxies in the left and
right panels respectively.  The control sample is presented in
red. The crosses mark the sample {\it mean} colors and their 1$\sigma$
errors along each axis (values are given in Table~\ref{sdsscoltable}).
The optical color offset between the variable and non-variable samples
are significant at the $\sim2\sigma$ level. This trend for variable
sources to become bluer during an outburst has been seen before (e.g.,
Trevese \& Vagnetti 2002). Note also that the the quasars are
generally bluer (i.e., lower u$-$r and r$-$z colors) than the
galaxies.}
\label{sdssCol}
\end{figure*}

We have established that there is a significant difference between the
variable and non-variable radio samples in terms of their optical
counterparts. The former have a higher matching rate with optical
identifications (given a fixed detection limit), which is at least in
part ascribed to the higher fraction of quasars.  We have already seen
that the magnitude distributions are not significantly different
(Fig.~\ref{apmids}), however, this does not rule out that significant
optical color differences may be present between the samples. The
clearest difference between galaxies and quasars (at least in the
restframe) is provided by the (u$-$r) vs (r$-$z) color plot (see
Fig~13 in Stoughton et al. 2002 for intermediate color-color plots).
This allows for sampling of the 4000\AA\ break in galaxies up to a
redshift of about unity. This break will offset the galaxies from the
quasars, which usually exhibit fairly flat, powerlaw continua (see
Fig.~3 in de Vries et al. 2003).  Our data are presented in
Fig.~\ref{sdssCol}. The sample means and their 1$\sigma$ errors
(indicated by the crosses in Fig.~\ref{sdssCol}) are presented in
Table~\ref{sdsscoltable}.  Both the variable quasar and galaxy
distributions in the color-color plane are (statistically) distinct
from the non-variable control sample. This trend for both the galaxies
and quasars to become bluer when they vary has been seen before (e.g.,
Giveon et al. 1999; Trevese \& Vagnetti 2002; Vagnetti et al. 2003; de
Vries et al. 2003).  Still, the color spread among the sources in the
sample is much larger than the {\it individual} color change as a
source goes through an outburst. Indeed, the studies of Giveon,
Trevese, Vagnetti and their respective collaborators all rely on
following individual objects through their outbursts, instead of using
a statistical sample.

The presence of the mean optical color offsets in our radio variable
sample implies two things: 1) on average, radio variable sources also
exhibit optical (color) variations, and 2) the optical variation
time-scales are shorter than our 7 year base-line, since the SDSS data
were not taken at the epoch of radio outburst. However, optical colors
for an individual source, while useful in differentiating between
quasars and galaxies, do not provide an effective selection mechanism
for (radio) variability. Nonetheless, as we have seen in the previous
sections, the presence of radio variability has a clear impact on the
sample's morphological makeup.

\section{Summary}

We have created a sample of 123 variable radio sources using two epoch
observations of a zero-dec strip toward the south Galactic cap.  This
sample spans the range of radio flux densities from $\sim2$ to 1000
mJy. It presents both in size and radio flux density coverage a unique
starting point for variability studies of more normal, less
AGN-dominated galaxies, especially toward the lower flux density
limits. We compared both our variable and non-variable samples to the
Sloan Digital Sky Survey optical data.

We found that the quasar fraction of the sample sharply declines as a
function of declining radio flux density levels. This is consistent
with earlier findings that the radio source population demographics
change as one samples at progressively lower flux density levels:
AGN-dominated systems tend to be found at the brighter radio flux
density levels ($> 10$ mJy), whereas star-forming and normal galaxies
dominate the counts at sub-mJy flux density levels (at 1.4 GHz,
e.g.,  Windhorst 2003).  Our variable sample contains a consistently
higher fraction of quasars than the non-variable control sample,
independent of radio flux density. While this explains part of our
almost $2\times$ higher optical matching rate of the variable sample
compared to the non-variable one (quasars are easier to detect at a
given brightness limit than galaxies), it does imply that our variable
sample contains on average slightly brighter sources (though not
significantly so, see Sect.~\ref{apmid}). Based on relative number
statistics, we estimate that quasars are about 5 times more likely to
harbor a variable radio source than galaxies. However, at flux density
levels $< 20$ mJy, the majority of (radio) variable sources are
identified as galaxies. And finally, galaxies and quasars that harbor
a variable radio source exhibit, on average, bluer optical colors than
hosts of non-variable sources.

All of this underlines the fact that both galaxies and quasars can
harbor variable radio sources, albeit at different occurrence rates.
Some of this is obviously due to the beamed nature of the (variable)
quasars, enhancing the variability both by boosting their brightnesses
and shortening the variability timescales\footnote{Even though blazars
represent a very small fraction of AGN, they are much more common in
variable samples.}. Nonetheless, especially toward lower radio flux
density limits, a statistically significant study of variability is
possible provided one starts with large enough samples. Large scale
optical surveys like the SDSS provide the crucial radio source optical
``environmental'' information. As more SDSS data become available on
the zero-dec strip, we will investigate this further.

\acknowledgements

The authors like to thank Steve Croft for useful discussions and
careful reading of this paper. WDVs work was performed under the
auspices of the U.S. Department of Energy, National Nuclear Security
Administration by the University of California, Lawrence Livermore
National Laboratory under contract No. W-7405-Eng-48. The authors also
acknowledge support from the National Radio Astronomy Observatory, the
National Science Foundations (grants AST 00-98259 and AST 00-98355),
the Space Telescope Science Institute, and Microsoft. This research
has made use of the NASA/IPAC extragalactic database (NED) which is
operated by the Jet Propulsion Laboratory, Caltech, under contract
with the National Aeronautics and Space Administration.


\begin{figure*}
\epsscale{2.0} 
\plotone{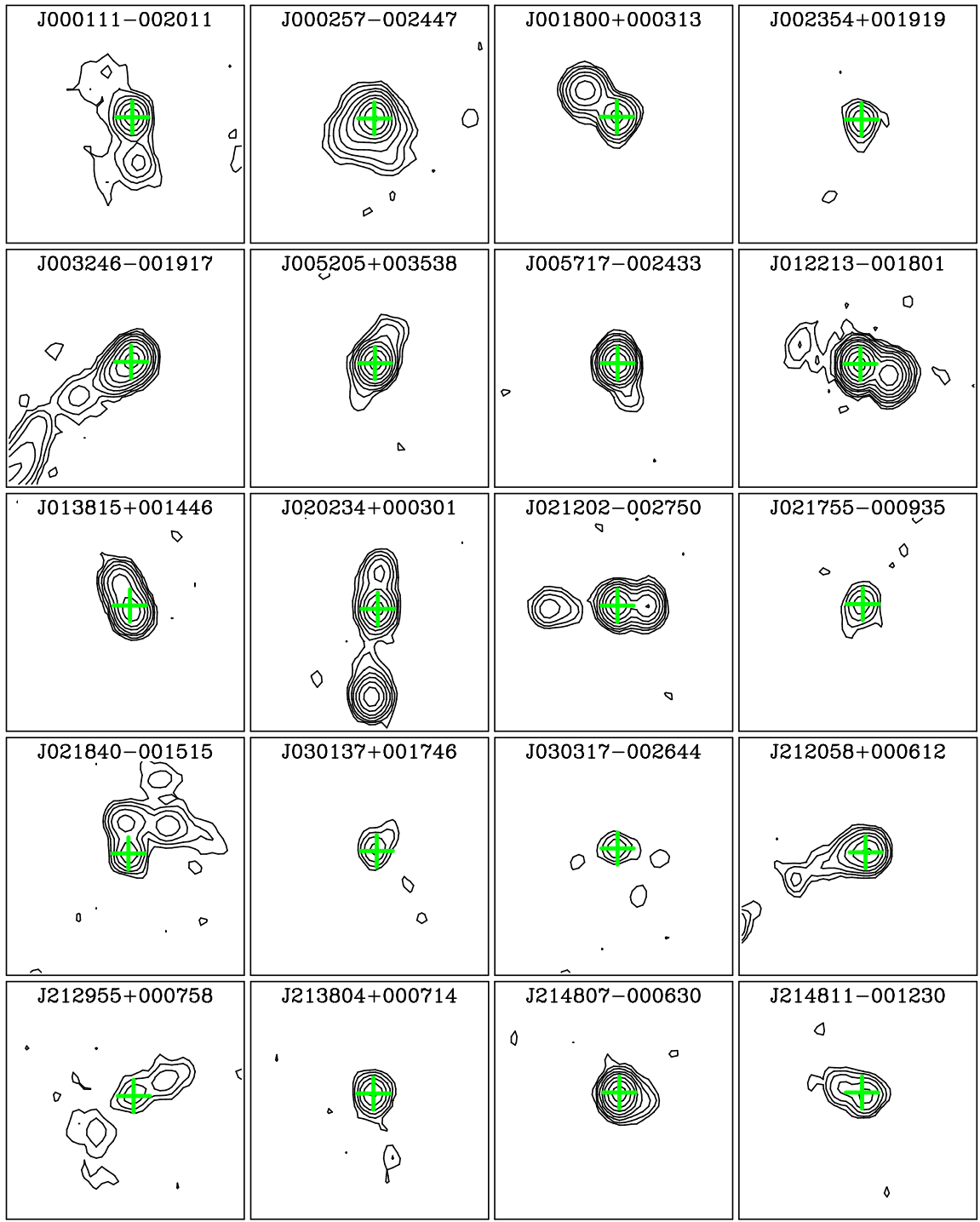}
\caption{FZDA images of the extended subset of our variable sample (28 sources in
total). The boxes are 72\arcsec$\times$72\arcsec\ in size. The variable component 
is indicated by the green cross.}
\label{radiomaps}
\end{figure*}

\begin{figure*}
\figurenum{10}
\epsscale{2.0} 
\plotone{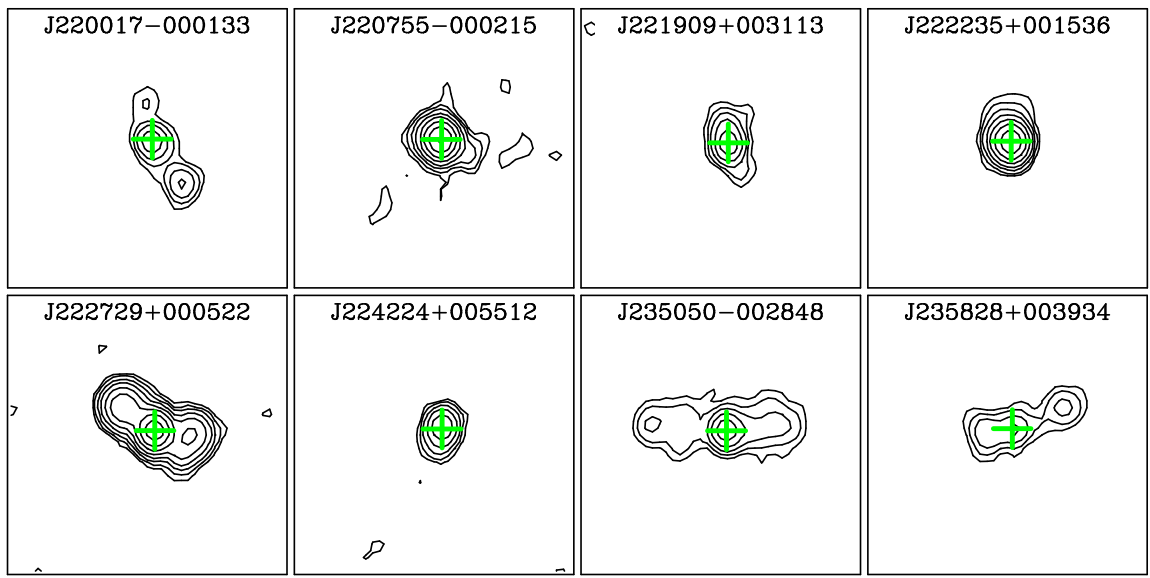}
\caption{continued}
\end{figure*}


\begin{deluxetable}{lrr}
\tablewidth{3.5in}
\tablecaption{FZDA / FZDB sample statistics\label{samples}}
\tablehead{\colhead{Sample} & \colhead{Total Number} & \colhead{Point Sources Only}}
\startdata
FZDA            & 7343 & 5550 \\
FZDB            & 9086 & 7286 \\
Combined        & 6605 & 5172 \\
FZDB not in A   & 2481 & 1846 \\
\enddata
\end{deluxetable}

\begin{deluxetable}{lrrrrcrrrrr}
\small
\tablewidth{7in}
\tablecaption{FIRST Zero-Dec Variable Sample\label{VarSample}}
\tablehead{
  \colhead{Source /} & 
  \colhead{RA (J2000)} & \colhead{Dec (J2000)} & 
  \colhead{  $F_{\footnotesize 2002}$} & 
  \colhead{  $F_{\footnotesize 1995}$} & 
  \colhead{FR\tablenotemark{a}} & 
  \colhead{S\tablenotemark{b}} &
  \colhead{z\tablenotemark{c}} & \colhead{r\tablenotemark{c}} & \colhead{ID\tablenotemark{c}} &
  \colhead{RM\tablenotemark{d}} 
  \\
  \colhead{Component} & \multicolumn{2}{c}{(core component)} & 
  \colhead{[mJy]} & \colhead{[mJy]} & 
  \colhead{} & \colhead{} & 
  \colhead{} & \colhead{} & \colhead{} & \colhead{}  
}
\startdata
J000111$-$002011 & 00 01 11.19 & $-$00 20 11.6 &   33.57 &   23.28 & 1.46 & 15.6 & 0.518 & 18.56 & Q& CL \\
J000257$-$002447 & 00 02 57.19 & $-$00 24 47.4 &  112.93 &  124.90 & 0.92 &  4.2 &       & 19.52 & G& CH \\
J000549$+$005048 & 00 05 49.92 &    00 50 48.1 &    8.87 &    5.13 & 1.77 & 10.1 &       & 21.59 & G& PS \\
J001158$-$000208 & 00 11 58.81 & $-$00 02 08.2 &    5.95 &   27.34 & 0.22 & 43.2 &       & 20.98 & Q& PS \\
J001507$-$000801 & 00 15 07.01 & $-$00 08 01.8 &   10.54 &   12.67 & 0.85 &  4.3 & 1.704 & 18.23 & Q& PS \\
J001611$-$001512 & 00 16 11.07 & $-$00 15 12.3 &  827.61 & 1050.26 & 0.80 & 11.2 & 1.575\tablenotemark{e} & 19.70 & Q& PS \\
J002354$+$001919 & 00 23 54.58 &    00 19 19.5 &    4.75 &    7.46 & 0.66 &  5.2 &       & 22.69 & G& CJ \\
J002738$+$000627 & 00 27 38.25 &    00 06 27.6 &    4.56 &    6.91 & 0.68 &  5.4 &       & 22.02 & Q& PS \\
J003007$-$000007 & 00 30 07.90 & $-$00 00 07.4 &   50.52 &   39.32 & 1.30 & 12.1 &       & 19.55 & Q& PS \\
J003127$+$003959 & 00 31 27.92 &    00 39 59.4 &    9.58 &   12.31 & 0.79 &  6.3 &       & 21.80 & Q& PS \\
J003536$-$000627 & 00 35 36.30 & $-$00 06 27.2 &   17.28 &   14.52 & 1.21 &  6.1 &       & 18.94 & G& PS \\
J003540$-$002529 & 00 35 40.22 & $-$00 25 29.2 &    4.01 &    6.57 & 0.63 &  6.4 &       & 21.82 & G& PS \\
J003703$+$003537 & 00 37 03.36 &    00 35 37.4 &   12.84 &   15.28 & 0.86 &  4.5 &       & 21.51 & Q& PS \\
J004332$+$002459 & 00 43 32.71 &    00 24 59.8 &  120.94 &  108.53 & 1.13 &  6.1 & 1.127\tablenotemark{e} & 19.20 & Q& PS \\
J004819$+$001457 & 00 48 19.11 &    00 14 57.1 &   98.76 &   89.40 & 1.12 &  5.6 & 1.536\tablenotemark{e} & 19.92 & Q& PS \\
J005205$+$003538 & 00 52 05.58 &    00 35 38.2 &   34.58 &   81.46 & 0.43 & 38.1 & 0.399 & 16.09 & Q& CJ \\
J005212$+$000945 & 00 52 12.47 &    00 09 45.2 &   11.92 &    9.77 & 1.24 &  5.5 &       &       &  & PS \\
J005225$+$002627 & 00 52 25.67 &    00 26 28.0 &   12.13 &    9.98 & 1.24 &  5.2 &       & 20.75 & G& PS \\
J005717$-$002433 & 00 57 17.01 & $-$00 24 33.2 &   89.47 &  114.74 & 0.79 & 11.6 & 2.790\tablenotemark{e} & 19.19 & Q& CJ \\
J010525$+$001121 & 01 05 25.52 &    00 11 21.7 &   33.97 &   41.73 & 0.83 &  8.3 &       & 21.45 & G& PS \\
J010745$+$003952 & 01 07 45.21 &    00 39 52.8 &   10.50 &    8.88 & 1.21 &  4.5 &       &       &  & PS \\
J010838$+$002814 & 01 08 38.56 &    00 28 14.6 &    5.33 &    3.91 & 1.40 &  4.2 &       &       &  & PS \\
J011106$+$000846 & 01 11 06.79 &    00 08 46.1 &    5.25 &    3.43 & 1.57 &  4.9 &       & 22.70 & Q& PS \\
J011515$+$001248 & 01 15 15.78 &    00 12 48.5 &   46.84 &   43.13 & 1.10 &  4.4 & 0.045 & 14.39 & G& PS \\
J012213$-$001801 & 01 22 13.92 & $-$00 18 01.0 &  386.60 &  331.60 & 1.18 &  8.3 &       & 20.23 & Q& CL \\
J012528$-$000555 & 01 25 28.85 & $-$00 05 55.8 & 1333.76 & 1481.35 & 0.91 &  4.8 & 1.076 & 16.47 & Q& PS \\
J012753$+$002516\tablenotemark{f} & 01 27 53.70 &    00 25 16.5 &  131.32 &   90.08 & 1.48 & 19.1 &       & 20.76 & Q& PS \\
J013457$+$003942 & 01 34 57.42 &    00 39 43.0 &    6.09 &    2.87 & 2.18 &  8.7 &       & 22.04 & Q& PS \\
J013815$+$001446 & 01 38 15.02 &    00 14 46.5 &   42.98 &   51.76 & 0.84 &  8.1 &       & 21.64 & G& CJ \\
J015329$-$002214 & 01 53 29.75 & $-$00 22 14.3 &   17.46 &   14.98 & 1.19 &  5.8 &       & 19.07 & Q& PS \\
J015528$+$001204 & 01 55 28.47 &    00 12 04.6 &   17.23 &   19.82 & 0.88 &  4.8 &       & 22.08 & G& PS \\
J015950$-$002407 & 01 59 50.09 & $-$00 24 07.2 &   10.68 &   12.63 & 0.86 &  4.3 &       & 21.50 & G& PS \\
J020141$+$003825 & 02 01 41.04 &    00 38 25.5 &    4.50 &    6.35 & 0.73 &  4.7 &       &       &  & PS \\
J020214$-$001748 & 02 02 14.30 & $-$00 17 48.3 &   75.12 &   60.00 & 1.27 & 11.5 &       & 21.36 & G& PS \\
J020234$+$000301 & 02 02 34.32 &    00 03 01.7 &   30.54 &   39.41 & 0.79 & 10.5 &       & 18.42 & G& CL \\
J020928$-$001224 & 02 09 28.85 & $-$00 12 24.9 &    4.10 &    2.51 & 1.69 &  4.2 & 0.152 & 16.02 & G& PS \\
J021202$-$002750 & 02 12 02.13 & $-$00 27 50.1 &   45.14 &   52.84 & 0.87 &  6.6 &       &       &  & CL \\
J021301$-$001815 & 02 13 01.13 & $-$00 18 15.0 &   41.55 &   48.44 & 0.87 &  6.5 &       & 21.98 & G& PS \\
J021553$+$001826 & 02 15 53.65 &    00 18 26.9 &   30.90 &   35.82 & 0.88 &  5.6 &       & 19.51 & G& PS \\
J021755$-$000935 & 02 17 55.99 & $-$00 09 35.9 &    3.96 &    5.85 & 0.70 &  4.4 &       & 19.86 & G& CJ \\
J021840$-$001515 & 02 18 40.55 & $-$00 15 15.9 &   10.06 &   13.17 & 0.78 &  6.4 & 1.171 & 18.81 & Q& CX \\
J022624$+$000746 & 02 26 24.61 &    00 07 46.2 &   43.84 &   48.74 & 0.91 &  4.4 &       & 22.56 & G& PS \\
J023105$+$000843 & 02 31 05.59 &    00 08 43.5 &   47.60 &   53.86 & 0.90 &  5.0 & 1.338 & 20.00 & Q& PS \\
J025321$+$000559 & 02 53 21.04 &    00 05 59.9 &  151.89 &   97.64 & 1.57 & 22.0 &       & 22.26 & G& PS \\
J025333$+$002431 & 02 53 33.66 &    00 24 31.7 &   16.86 &   14.86 & 1.15 &  4.6 &       &       &  & PS \\
J025404$-$002628 & 02 54 04.60 & $-$00 26 28.7 &    4.39 &    2.71 & 1.67 &  5.2 &       & 21.97 & G& PS \\
J025859$+$003618 & 02 58 59.65 &    00 36 18.3 &   29.45 &   23.75 & 1.26 &  9.4 &       & 22.28 & Q& PS \\
J025928$-$002000\tablenotemark{f} & 02 59 28.52 & $-$00 20 00.1 &  243.47 &  225.85 & 1.09 &  4.3 & 2.001 & 17.34 & Q& PS \\
J030137$+$001746 & 03 01 37.58 &    00 17 46.1 &    6.86 &    4.45 & 1.58 &  7.0 &       & 21.47 & G& CJ \\
J030317$-$002644 & 03 03 17.01 & $-$00 26 45.0 &    7.41 &    3.68 & 2.06 &  4.9 &       &       &  & CJ \\
J030702$+$000651 & 03 07 02.04 &    00 06 51.9 &    5.84 &    8.08 & 0.74 &  5.4 &       &       &  & PS \\
J030834$+$003303 & 03 08 34.31 &    00 33 03.7 &    4.16 &    1.83 & 2.35 &  6.4 & 0.031 & 14.91 & G& PS \\
J030933$-$001901 & 03 09 33.27 & $-$00 19 01.2 &    6.31 &    4.58 & 1.41 &  4.6 &       &       &  & PS \\
J031006$+$001549 & 03 10 06.69 &    00 15 49.9 &    1.36 &    5.55 & 0.26 & 11.1 & 0.109 & 18.54 & G& PS \\
J031118$+$000816 & 03 11 18.51 &    00 08 16.4 &   29.30 &   36.54 & 0.81 &  9.1 &       &       &  & PS \\
J031202$-$000442 & 03 12 02.50 & $-$00 04 42.5 &    6.42 &    8.59 & 0.77 &  5.2 & 0.038 & 13.50 & G& PS \\
J031345$-$000720 & 03 13 45.04 & $-$00 07 20.3 &    3.63 &    6.06 & 0.62 &  6.1 & 2.519 & 20.04 & Q& PS \\
J031353$-$000004 & 03 13 53.47 & $-$00 00 04.1 &   10.13 &    7.03 & 1.47 &  7.7 &       & 21.16 & G& PS \\
J031357$+$003506 & 03 13 57.10 &    00 35 06.9 &   16.54 &   20.23 & 0.83 &  6.7 &       &       &  & PS \\
J031452$+$001346 & 03 14 52.08 &    00 13 46.5 &    6.77 &   10.05 & 0.69 &  7.3 & 3.202 & 19.54 & Q& PS \\
J031609$+$000107 & 03 16 09.55 &    00 01 07.9 &    4.02 &    6.03 & 0.69 &  4.6 &       &       &  & PS \\
J031634$-$002039 & 03 16 34.96 & $-$00 20 39.6 &   18.17 &   15.63 & 1.18 &  5.5 &       &       &  & PS \\
J031814$-$002948\tablenotemark{f} & 03 18 14.43 & $-$00 29 48.9 &  113.20 &   93.91 & 1.22 &  9.7 &       & 21.77 & G& PS \\
J032007$+$000354 & 03 20 07.01 &    00 03 54.0 &   12.17 &   10.20 & 1.22 &  5.0 &       & 22.10 & G& PS \\
J212000$-$001159 & 21 20 00.72 & $-$00 11 59.5 &   42.94 &   48.33 & 0.90 &  4.9 &       &       &  & PS \\
J212447$+$000747 & 21 24 47.34 &    00 07 47.6 &    5.79 &    3.82 & 1.56 &  5.3 &       &       &  & PS \\
J213206$+$003520 & 21 32 06.15 &    00 35 20.1 &   13.65 &   11.95 & 1.16 &  4.1 &       &       &  & PS \\
J213638$+$004154\tablenotemark{f} & 21 36 38.57 &    00 41 54.3 & 4136.68 & 3546.71 & 1.18 &  8.3 & 1.932\tablenotemark{e} & 16.79 & q & PS \\
J213748$+$001219 & 21 37 48.43 &    00 12 19.9 &   41.69 &   36.02 & 1.17 &  7.0 & 1.666\tablenotemark{e} & 17.92 & q& PS \\
J213804$+$000714 & 21 38 04.06 &    00 07 14.7 &   10.31 &   16.33 & 0.64 & 11.6 &       &       &  & CJ \\
J214138$+$000319 & 21 41 38.55 &    00 03 19.8 &   11.08 &   14.37 & 0.79 &  6.8 &       &       &  & PS \\
J214324$+$003502 & 21 43 24.37 &    00 35 02.7 &   36.03 &   45.22 & 0.81 &  9.7 & 2.030\tablenotemark{e} & 19.37 & q& PS \\
J214419$+$002055 & 21 44 19.88 &    00 20 55.8 &   10.34 &    7.31 & 1.44 &  7.5 &       &       &  & PS \\
J214613$+$000930 & 21 46 13.31 &    00 09 30.8 &    9.94 &    7.88 & 1.29 &  5.6 &       &       &  & PS \\
J214807$-$000630 & 21 48 07.73 & $-$00 06 30.6 &   68.57 &   90.18 & 0.77 & 12.7 &       &       &  & CJ \\
J214811$-$001230 & 21 48 11.48 & $-$00 12 30.6 &    7.76 &    9.95 & 0.80 &  4.7 &       &       &  & CJ \\
J215349$+$003119 & 21 53 49.75 &    00 31 19.6 &   82.50 &  103.65 & 0.81 & 10.4 &       &       &  & PS \\
J215353$-$001339 & 21 53 53.89 & $-$00 13 39.5 &   13.48 &   17.27 & 0.79 &  7.3 &       &       &  & PS \\
J215359$+$004412 & 21 53 59.83 &    00 44 12.6 &    9.76 &    6.73 & 1.48 &  7.8 & 1.030\tablenotemark{e} & 19.07 & q& PS \\
J215733$-$000340\tablenotemark{f} & 21 57 33.66 & $-$00 03 40.5 &    5.36 &    3.60 & 1.53 &  4.9 &       &       &  & PS \\
J215949$+$005146 & 21 59 49.91 &    00 51 46.7 &   12.29 &    9.21 & 1.36 &  7.8 &       &       &  & PS \\
J215954$-$002150 & 21 59 54.46 & $-$00 21 50.1 &    4.07 &    2.46 & 1.71 &  4.7 & 1.960\tablenotemark{e} & 16.98 & q& PS \\
J220017$-$000133 & 22 00 17.37 & $-$00 01 33.6 &    7.84 &    6.15 & 1.30 &  4.9 &       &       &  & CL \\
J220755$-$000215 & 22 07 55.25 & $-$00 02 15.0 &   78.40 &   61.93 & 1.28 & 11.9 &       &       &  & CJ \\
J220822$+$002352 & 22 08 22.88 &    00 23 52.7 &    4.62 &    2.42 & 1.97 &  6.1 &       &       &  & PS \\
J221001$-$001309\tablenotemark{f} & 22 10 01.82 & $-$00 13 09.9 &  125.83 &  115.31 & 1.10 &  4.8 &       &       &  & PS \\
J221031$-$001356 & 22 10 31.46 & $-$00 13 56.1 &   14.06 &   12.00 & 1.19 &  5.2 &       &       &  & PS \\
J221909$+$003113 & 22 19 09.40 &    00 31 13.4 &    9.13 &   11.70 & 0.80 &  5.3 &       &       &  & CJ \\
J222036$+$003334 & 22 20 36.33 &    00 33 34.2 &   15.27 &   12.76 & 1.22 &  5.8 &       &       &  & PS \\
J222135$-$001100 & 22 21 35.00 & $-$00 11 00.1 &   14.11 &    8.64 & 1.66 & 12.6 &       &       &  & PS \\
J222235$+$001536 & 22 22 35.87 &    00 15 36.7 &   46.04 &   51.13 & 0.91 &  4.5 &       &       &  & CJ \\
J222704$+$004517 & 22 27 04.24 &    00 45 17.5 &    8.13 &    5.57 & 1.49 &  7.0 &       &       &  & PS \\
J222726$+$001059 & 22 27 26.53 &    00 10 59.3 &    6.05 &    4.52 & 1.37 &  4.2 &       &       &  & PS \\
J222729$+$000522 & 22 27 29.08 &    00 05 22.2 &   77.84 &   91.64 & 0.86 &  7.4 & 1.510\tablenotemark{e} & 18.86 & q& CL \\
J222744$+$003450 & 22 27 44.58 &    00 34 50.6 &   22.13 &   30.53 & 0.74 & 12.1 & 1.540\tablenotemark{e} & 19.14 & q& PS \\
J222758$+$003705 & 22 27 58.13 &    00 37 05.2 &   68.69 &   99.12 & 0.70 & 17.3 &       &       &  & PS \\
J223047$+$002756 & 22 30 47.46 &    00 27 56.3 &    8.54 &   13.82 & 0.63 & 11.4 &       &       &  & PS \\
J224224$+$005512 & 22 42 24.14 &    00 55 13.0 &   17.71 &   13.44 & 1.34 &  9.5 &       &       &  & CJ \\
J224331$-$001233 & 22 43 31.94 & $-$00 12 33.1 &   13.85 &   21.40 & 0.66 & 14.8 & 2.040\tablenotemark{e} & 18.45 & q& PS \\
J224448$-$000619 & 22 44 48.11 & $-$00 06 19.8 &    8.24 &    5.58 & 1.51 &  7.3 &       &       &  & PS \\
J224627$-$001214\tablenotemark{f} & 22 46 27.68 & $-$00 12 14.2 &   85.55 &   56.00 & 1.55 & 20.9 &       &       &  & PS \\
J224730$+$000006\tablenotemark{f} & 22 47 30.19 &    00 00 06.1 &  464.52 &  183.71 & 2.56 & 43.8 & 0.094\tablenotemark{e} & 18.50 & q& PS \\
J224922$+$001804 & 22 49 22.28 &    00 18 04.6 &   10.41 &    8.51 & 1.25 &  5.3 &       &       &  & PS \\
J225852$-$001857 & 22 58 52.94 & $-$00 18 57.3 &    3.45 &    1.28 & 2.80 &  6.0 &       &       &  & PS \\
J230157$+$000351 & 23 01 57.85 &    00 03 52.0 &    3.75 &    7.04 & 0.55 &  7.8 &       & 23.23 & Q& PS \\
J230314$+$000052 & 23 03 14.85 &    00 00 52.2 &    3.72 &    1.72 & 2.24 &  5.8 &       &       &  & PS \\
J230655$+$003638 & 23 06 55.16 &    00 36 38.1 &   12.99 &   15.19 & 0.87 &  4.3 &       &       &  & PS \\
J231541$+$002936 & 23 15 41.67 &    00 29 36.6 &   21.93 &   13.94 & 1.60 & 15.7 &       & 20.81 & Q& PS \\
J231558$-$001205 & 23 15 58.64 & $-$00 12 05.5 &    3.52 &    6.27 & 0.58 &  7.0 &       & 22.85 & G& PS \\
J231845$-$000754 & 23 18 45.81 & $-$00 07 54.7 &    3.03 &    5.34 & 0.59 &  5.5 & 0.867 & 19.44 & Q& PS \\
J231856$+$001437\tablenotemark{g} & 23 18 56.66 &    00 14 37.7 &   20.16 &   18.34 & 1.12 &  4.2 & 0.030   & 12.76 & G& PS \\
J231910$+$001859 & 23 19 10.33 &    00 18 59.0 &   28.90 &   33.36 & 0.88 &  5.6 &       & 22.60 & Q& PS \\
J232038$+$003139 & 23 20 38.01 &    00 31 39.8 &   72.69 &   82.95 & 0.89 &  5.7 & 1.911 & 19.03 & Q& PS \\
J232323$+$003327 & 23 23 23.95 &    00 33 27.5 &   16.73 &   12.52 & 1.36 &  9.2 &       &       &  & PS \\
J233448$-$001400 & 23 34 48.06 & $-$00 14 01.0 &   24.85 &   29.64 & 0.85 &  6.8 &       & 21.56 & Q& PS \\
J233822$+$001146 & 23 38 22.35 &    00 11 46.6 &   18.19 &   15.44 & 1.20 &  6.0 &       & 22.24 & Q& PS \\
J233852$+$004843 & 23 38 52.46 &    00 48 43.5 &   11.32 &    9.18 & 1.26 &  5.9 &       &       &  & PS \\
J234623$+$004301 & 23 46 23.73 &    00 43 01.1 &   13.49 &   11.79 & 1.17 &  4.6 & 2.861 & 19.09 & Q& PS \\
J234939$-$001315 & 23 49 39.90 & $-$00 13 15.2 &    8.46 &    6.97 & 1.24 &  4.2 & 1.267 & 20.22 & Q& PS \\
J235022$+$001232 & 23 50 22.41 &    00 12 32.4 &    3.92 &    2.03 & 2.00 &  5.1 &       &       &  & PS \\
J235050$-$002848 & 23 50 50.72 & $-$00 28 48.7 &   13.12 &   11.57 & 1.15 &  4.1 &       & 19.62 & G& CL \\
J235409$-$001948\tablenotemark{f} & 23 54 09.18 & $-$00 19 48.1 &  384.58 &  344.79 & 1.13 &  6.1 & 0.462 & 17.93 & Q& PS \\
J235823$+$000213 & 23 58 23.91 &    00 02 13.2 &    8.03 &   10.67 & 0.77 &  5.7 &       &       &  & PS \\
\tableline
J001800$+$000313\tablenotemark{h} & 00 18 00.79 &    00 03 17.9 &   20.76 &   18.82 & 1.12 &  4.2 &       &       &  & HS \\
J003246$-$001917\tablenotemark{h} & 00 32 46.02 & $-$00 19 17.8 &   81.29 &   60.80 & 1.35 & 14.5 &       &       &  & HS \\
J212058$+$000612\tablenotemark{h} & 21 20 59.00 &    00 06 12.7 &   31.17 &   28.05 & 1.13 &  5.3 &       &       &  & HS \\
J212955$+$000758\tablenotemark{h} & 21 29 55.68 &    00 07 59.0 &    1.17 &    3.03 & 0.42 &  4.5 &       &       &  & CL \\
J235828$+$003934\tablenotemark{h} & 23 58 28.77 &    00 39 34.1 &    2.48 &    4.16 & 0.62 &  4.3 &       &       &  & CL \\
\enddata
\vspace{-6mm}\tablenotetext{a}{Corrected flux density ratio, see Eqn.~\ref{vreqn2}.\vspace{-4mm}} 
\tablenotetext{b}{Significance (in units of $\sigma$ rms) as defined
by: $S = \left| \frac{\mbox{FR}-1}{\chi\left(\mbox{FR}+1\right)}
\right|$, using the $\chi$ definition from Eqn. 4.\vspace{-2mm}}
\tablenotetext{c}{SDSS redshift, r-band magnitude, and optical
identification, except where noted. Sources in the list with RA between 
21$^{\rm h}$ and 23$^{\rm h}$ are not covered by SDSS DR1.\vspace{-4mm}}
\tablenotetext{d}{Radio morphology: PS=pointsource, CJ=core-jet, CL=core + lobes, CX=complex, 
CH=core + halo, and HS=possible hotspot.\vspace{-5mm}}
\tablenotetext{e}{Redshift taken from literature. If the ID is in lower case then the
optical data are from the literature as well.\vspace{-4mm}}
\tablenotetext{f}{Also meets the FZDB-NVSS variability criterion. Only
sources brighter in the FZDB epoch were considered.\vspace{-4mm}}
\tablenotetext{g}{NGC 7603}
\tablenotetext{h}{Variability of this source is spurious (see Sect.~\ref{radiomorph})}
\normalsize
\end{deluxetable}

\begin{deluxetable}{lcc}
\tablewidth{2.6in}
\tablecaption{APM match statistics, $3\arcsec$ search radius\label{apmmatch}}
\tablehead{\colhead{Sample\tablenotemark{a}} & \colhead{Detected} & \colhead{Rate}}
\startdata
Variable & 66 & 53.7\%\\
\tableline
Control 1  &  17 & 13.8\%\\
Control 2  &  29 & 23.6\%\\
Control 3  &  21 & 17.1\%\\
Control 4  &  21 & 17.1\%\\
Control 5  &  20 & 16.3\%\\
Control 6  &  27 & 22.0\%\\
Control 7  &  23 & 18.7\%\\
Control 8  &  17 & 13.8\%\\
Control 9  &  21 & 17.1\%\\
Control 10 &  22 & 17.9\%\\
Control 11 &  23 & 18.7\%\\
Control 12 &  21 & 17.1\%\\
\tableline
Mean Control & 21.8$\pm$3.5 & 17.7$\pm$2.8\% \\
\enddata
\tablenotetext{a}{All samples contain 123 sources}
\end{deluxetable}

\begin{deluxetable}{lcccc}
\tablewidth{4.2in}
\tablecaption{SDSS match statistics, $3\arcsec$ search radius\label{sdssmatch}}
\tablehead{\colhead{Sample\tablenotemark{a}} & \colhead{Detected} & \colhead{Rate} &
\colhead{Quasars\tablenotemark{b}} & \colhead{Galaxies\tablenotemark{b}}}
\startdata
Variable & 64 & 82.0\%& 53.1\% & 46.9\% \\
\tableline
Control 1  &  28 & 37.3\%&17.9\%&82.1\% \\
Control 2  &  30 & 40.0\%&13.3\%&86.7\% \\
Control 3  &  21 & 28.0\%&14.3\%&85.7\% \\
Control 4  &  36 & 48.0\%&13.9\%&86.1\% \\
Control 5  &  26 & 34.7\%&26.9\%&73.1\% \\
Control 6  &  32 & 42.7\%&18.8\%&81.2\% \\
Control 7  &  29 & 38.7\%&24.1\%&75.9\% \\
Control 8  &  33 & 44.0\%& 9.1\%&90.9\% \\
Control 9  &  23 & 30.7\%&21.7\%&78.3\% \\
Control 10 &  30 & 40.0\%&16.7\%&83.3\% \\
Control 11 &  31 & 41.3\%&29.0\%&71.0\% \\
Control 12 &  22 & 29.3\%&18.2\%&81.8\% \\
\tableline
Mean Control & 28.4$\pm$4.6 & 37.9$\pm$6.2\% & 18.7$\pm$5.9\%  & 81.3$\pm$5.9\%  \\
\enddata
\tablenotetext{a}{Only 75 out of 123 variable sources are covered by the
SDSS DR1 data. The rates have been corrected accordingly.}
\tablenotetext{b}{Relative fraction of quasars and galaxies, based on the
SDSS morphology.}
\end{deluxetable}

\begin{deluxetable}{rrr}
\tablewidth{3.0in}
\tablecaption{Mean Optical colors of Variable and Non-Variable Samples\label{sdsscoltable}}
\tablehead{\colhead{Sample} & \colhead{(U-R)} & \colhead{(R-Z)}}
\startdata
 {\bf Quasars} - variable & 0.89$\pm$0.13 & 0.40$\pm$0.07 \\
       non-variable & 1.37$\pm$0.17 & 0.60$\pm$0.08 \\
     nominal offset\tablenotemark{a} & $-$0.48       & $-$0.20 \\
     significance\tablenotemark{b} & 2.3 & 1.8\\
\tableline
{\bf Galaxies} - variable & 1.81$\pm$0.21 & 1.09$\pm$0.11 \\
       non-variable & 2.28$\pm$0.09 & 1.20$\pm$0.05 \\
     nominal offset\tablenotemark{a} & $-$0.47       & $-$0.11 \\
     significance\tablenotemark{b} & 2.0 & 0.9\\
\enddata
\tablenotetext{a}{Between non-variable and variable samples.}
\tablenotetext{b}{Defined as nominal offset divided by the rms of the uncertainties.}
\end{deluxetable}

\end{document}